\documentclass[runningheads]{llncs}
\usepackage[T1]{fontenc}
\usepackage{color}
\usepackage{graphicx,verbatim}
\begin{document}
\title{Foundational values for foundation models}
%\titlerunning{Abbreviated paper title}
% If the paper title is too long for the running head, you can set
% an abbreviated paper title here
%

\author{John S. H. Baxter\inst{1}\orcidID{0000-0003-3548-4343} \and
Elodie Germani\inst{1}\orcidID{0000-0002-5786-9538}}
\authorrunning{J.S.H. Baxter \& E. Germani}
% First names are abbreviated in the running head.
% If there are more than two authors, 'et al.' is used.
%
\institute{Université de Rennes, Inserm, Laboratory Traitement du Signal et de l'Image (LTSI - UMR 1099), F-35000 Rennes, France. \email{john.baxter@univ-rennes.fr}}

\maketitle              % typeset the header of the contribution
\begin{abstract}
Research values, properties with a distinctive normative dimension, often affect how technological research is performed in both direct and indirect ways by influencing how technical decisions are made. In machine learning for medical imaging, understanding these values can
be important for understanding why particular researchers justify the decisions made in their publications and explain why certain technologies become ubiquitous (or not) in the scientific literature and in the clinic. This article explores one of these technologies, \textit{foundation models}, finding detailed justifications both for their use and abstention from their use. By taking a Socratic approach to research values arising from this specific technical decision, this article aims to better illustrate how foundation models fit into the philosophy of machine learning in medicine.

\keywords{Research values \and machine learning \and foundation models}
\end{abstract}
\section{Introduction}
Although the perceived ideal form of scientific research is purely objective, determining which research questions are asked and how a research program is constructed depends heavily on the values espoused by researchers and the social, cultural, and financial contexts in which the research takes place \cite{duffy2009values}. For example, medical research ethics give explicit outlines for how different types of medical research should be performed in terms of participant recruitment, study design, constitution of a control group, etc… based on fundamentally ethical, rather than epistemological, principles \cite{masic2014ethics}. At an even higher level, the choice of a research topic itself is often dictated by the values espoused by governmental organisations and granting agencies which can sometimes be entirely separate from the potential for gaining knowledge within the scientific community \cite{xiang2025research}. 

But the influence of values in technological research extends much deeper than experimental design or problem selection, towards the very selection of techniques themselves. That is, research values dictate what the space of technical decisions is and which should be considered the ``best’’ above and beyond mere performance. Certain values in this area are uncontroversial, such as explicitly preferring smaller machine learning models to limit resource consumption, whereas others may be significantly moreso, such as choosing a particularly popular ``new’’ model to increase the probability of impactful publication.

The purpose of this article is to explore the research values that would motivate a particular choice in how artificial intelligence is used for medical image computing, specifically whether to use a foundation model or to train a model ``from scratch.’’

\section{What are foundation models?}
Briefly, \textit{foundation models} are pretrained architectures that are explicitly designed to handle multiple tasks with relatively minimal adaptation \cite{khan2025comprehensive}. The most widespread current example of a foundation model would be general pretrained transformers used as the technical backend for large-language models. These models are pretrained on a particular problem, such as filling in missing words or doing next-word prediction, with a large corpus of data that is not specific to an individual problem. If they were then evaluated on that same task, they would not be considered a foundation model, but instead they are applied to and evaluated on how well they can perform other tasks, such as recalling facts or performing logical reasoning. 

For medical imaging, these models come in a number of varieties that may or may not interface with language \cite{khan2025comprehensive}. They also tend to be more domain-specific, with foundation models specific to imaging modalities (such as MRI \cite{sun2025foundation} and CT \cite{blankemeier2024merlin}) or processing tasks (such as segmentation \cite{zhu2024medical} or registration \cite{tian2024unigradicon}). Common terms associated with the use of foundation models include notions of ``zero-shot'' learning a task is performed in application time for which a model has not been explicitly trained (hence it being a foundation model) \cite{rezaei2020zero} or ``model adaptation'' in which a model pretrained for one problem is slightly modified (such as through the addition of an additional ``head'') to solve a different task \cite{wang2023real}. These terms illustrate the different ways in which foundation models can be used after having been pretrained, but also indicate a certain question of \textit{degree} to this use; that is, one could imagine that as the adaptation becomes more complex, less of the underlying foundation model remains. This indicates that the question of whether to use foundation models is not binary, but rather exists along a continuous spectrum \cite{baxter2025exploring}.

\section{What are research values?}
\textit{Research values} are properties that a particular technique has that also posses some normative dimension used to define whether or not a particular experimental or technical decision is ``good.'' For example, one may consider a particular algorithm to be better than another because it is faster or consumes less memory, thus making ``speed'' and ``memory efficiency'' potential research values. However, these values do not have to be so direct but may instead affect technical decisions through other values as intermediaries. For example, if one espouses environmentalism as a value, then it may lead to also (albeit more subconsciously) espousing speed and memory efficiency, since faster and more memory efficient algorithms would have a lower carbon footprint. This leads to the inherent complexity of the relationship between research values and specific technical decisions; it can be unclear which values are instrumental, require mediation by an instrumental value, or are entirely standalone \cite{baxter2025exploring}.
Research values can also come in various flavours such as \textit{epistemic research values}, i.e. simplicity and reproducibility etc..., and \textit{non-epistemic research values}, i.e. societal/environmental impact, security, etc... although certain philosophers have debated the boundaries and utilities of these broad families \cite{rooney2017borderlands}. 

This article uses the Socratic methodology outlined by Baxter \& Eagleson \cite{baxter2025exploring}. This approach works in an iterative manner by doing the following:
\begin{enumerate}
    \item Take a given value (or technical decision) and hypothesize a reason (no matter how direct or indirect) a person may have that could motivate that value or justify that decision
    \item Determine if said connection is direct or if it is possible to have some intermediate explanation as to why the two values are connected
        \begin{itemize}
            \item In the latter case, determine what the intermediate explanation is and add that as another value, repeating this directness check with both previous values    
        \end{itemize}
    \item Determine if there is any immediate and direct conflict between the newly added value and the others already in the graph.
    \item After this is complete, the values with their supporting arrows can be put into the graph.
    \item Repeat until all values in the graph are either fundamental or have already been explored.
\end{enumerate}
This approach is often purely theoretical, although it can also be augmented by literature reviews that equate particular design decisions with particular properties, often more low-level direct ones such as speed, accuracy, memory consumption, etc... This approach helps to distinguish between instrumental values that explain how other, more fundamental values lead to particular technical decisions and thus how different sets of fundamental values could nevertheless lead to the same outcomes. This methodology has the benefit of side-stepping common problems in the philosophy of science regarding research values (such as the aforementioned \textit{epistemic/non-epistemic} distinction) in favour of an approach targeted to a very narrow technical question.

\section{A network of research values underlying the use of foundation models}
\begin{figure}[t]
    \centering
    \includegraphics[width=0.985\linewidth]{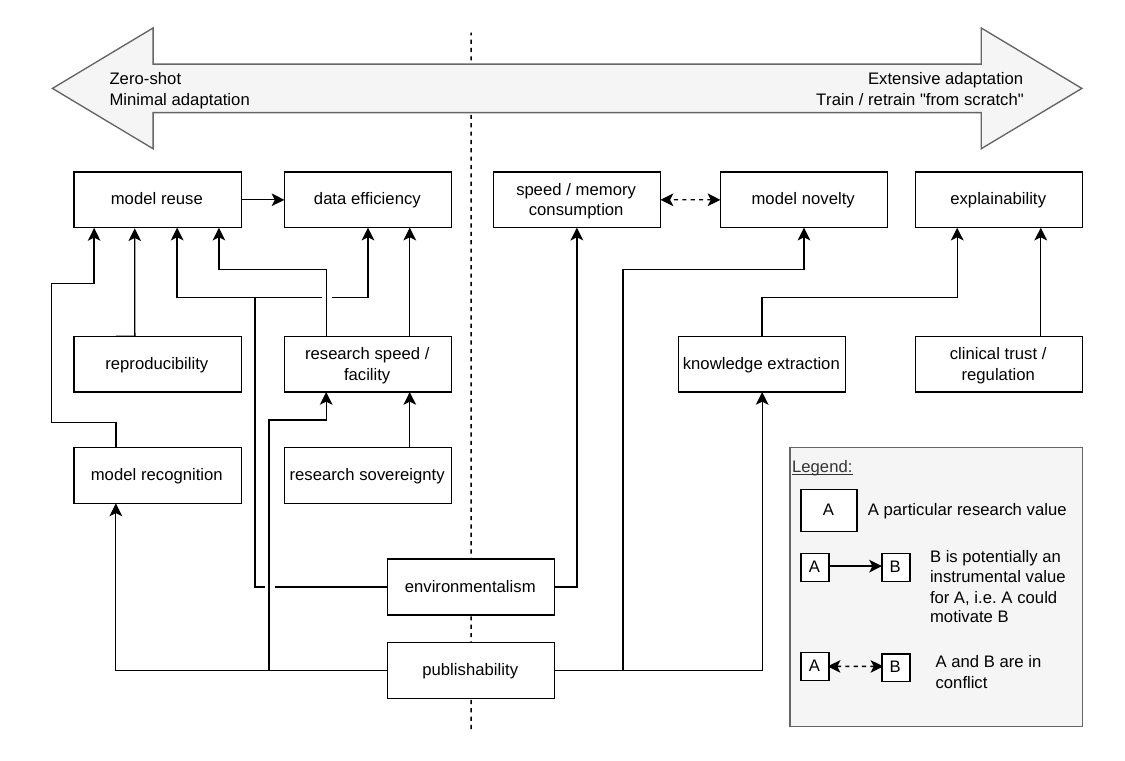}
    \vspace{-7mm}\caption{Graph of research values related to the degree of foundation model use.}
    \label{fig:network}
\end{figure}

The results of our Socratic investigation are shown in Figure \ref{fig:network}. Starting from the top of the network and moving downward, the justification for the position of the research value to the left (in favour of foundation model use with minimal adaptation) or to the right (in favour of more extensive adaptation or for constructing/training a model from scratch) are presented in the following subsections.

\subsection*{Data efficiency}
The primary stated motivation for the use of foundation models is \textit{data efficiency} \cite{zhang2026data} (or equivalently, the ability to overcome data scarcity), which means that researchers need fewer datasets to train a particular model, especially in the case where zero-shot or similar methodologies are used and there are few added components that need training \cite{khan2025comprehensive}. However, the specific number of datasets varies heavily based on the theoretical complexity of the problem (especially for tasks related to diagnostics and medical reasoning) and the quantity of annotation provided (i.e. less for diagnosis but significantly more for problems such as segmentation). 

\subsection*{Model reuse}
Immediately related to data efficiency is the ability to \textit{reuse a specific model}, which is the defining characteristic of foundation models. Since one can imagine being data efficient without reusing a model or being able to reuse a model without necessarily minimising the amount of data in the process, these values are conceptually separate. In practice, however, they are very much correlated. Model reuse provides the technical mechanism supporting data efficiency as pretraining on other tasks or more generic problems has been found to increase the accuracy or possibility of few-shot or zero-shot learning \cite{khan2025comprehensive}. Like data efficiency, model reuse is clearly an instrumental value forced upon any particular researcher by the constraints of their particular problem, rather than a fundamental value they would otherwise espouse for its own sake.

\subsection*{Speed and memory consumption}
What is rarely discussed in papers that use foundation models (rather than papers that propose new ones) is that pretrained foundation models tend to be significantly larger and heavier-duty than the majority of single-purpose models in the literature. This is sensible in that foundation models, by their vary conception, need to have the capacity to learn significantly more information about the world, as said information may be necessary for applying the model in a specific but yet unknown domain and because the pretraining datasets should be large enough to provide this level of information. By consequence, when a foundation model is used for a new specific task, a large portion of its architecture should go under-utilised. Note that this is different from the idea of \textit{dead neurons} in which certain elements in an architecture consistently have zero activation and thus cannot contribute to any computation \cite{jiang2022delve}; for a well-sized and well-trained foundation model, many neurons will have non-zero activations, but they simply will never contribute in any meaningful way to the final prediction on in a specific domain.  Thus, if one is concerned with the speed and memory consumption of a model, there are definite hurdles to the use of these large generically-pretrained foundation models and thus more of a reason to develop ones own model or engage in extensive retraining and adaptation steps to sparsify, isolate, and remove these elements i.e. model distillation \cite{liu2024wisdom}.

\subsection*{Model novelty}
\textit{Model novelty} or the ability for a model to be architecturally distinct in a meaningful way both supports the development of one's own models for obvious reasons. However, it also stands in contrast to speed and memory consumption, despite both generally supporting the right side of the foundation model use spectrum. This is because as machine learning has matured as a field, new models have been getting significantly more complex in terms of number of parameters and model depth \cite{villalobos2022machine}. Although some researchers do focus on novel ``frugal'' models, novelity in model architecture tends to involve additions to architectures rather than subtractions \cite{girdhar2025comprehensive}. Certain families of architectures (such as traditional CNN's for image classification) are now clearly outside of the purview of the technically novel. Despite this contradiction, there does not appear to be a possibility of an internal tension on the part of the researcher; it is hard to imagine someone consciously espousing both values to a very high degree simultaneously. In Baxter \& Eagleson's \cite{baxter2025exploring} original analysis, was reserved for pairs of values that support opposite ends of the spectrum defined by the technical question. In this case, however, they seem to both support the right side of the spectrum (training new architectures or extensive adaptation) mostly because they preclude the left (using foundation models with minimal adaptation). People who espouse speed and memory consumption avoid the computational cost of foundation model use by training simpler architectures and people who espouse model novelty avoid the lack of novelty in using other people's foundation models by training their own models with novel components.

\subsection*{Explainability}
Similarly to the previous points, foundation models suffer from a distinct lack of \textit{explainability} or the capability to intuitively understand the high-level functioning of a specific model. This is for several reasons: (i) the size and depth of pretrained foundation models limit the degree to which their thought-processes can be directly modelled and understood, (ii) the underlying models may be too unwieldy for traditional post-hoc explainability methods such as GradCAM \cite{selvaraju2020grad} or adversarial explanations \cite{ignatiev2019relating}, and (iii) a lack of information about model weight uncertainties (in pretraining and almost definitionally with respect to the new domain) needed for techniques such as Bayesian neural network methods \cite{arbel2026primer} that explain model uncertainty in a principled way. Alternatively, developing one's own models gives researchers extensive opportunities to address these concerns and thus to apply the theoretical infrastructure of explainability methods to their models. There are some movements towards addressing this specific to vision-language models \cite{bai2025evlf,nie2025explainable} which use the linguistic components of the foundation model to textually self-explain its process, although in other domains, there has been significant issues with the alignment of machine generated self-explanations with more direct computational techniques \cite{turpin2023language}. Outside of research, explainability tends to be an instrumental value, motivated by other more generic values such as trust, regulation, and fairness, although within the research community, this is more questionable as explainability has been such an active area of research in itself for almost a decade.

\subsection*{Reproducibility}
One of the canonical epistemic research values is \textit{reproducibility} \cite{rooney2017borderlands}, is the ability to do similar experiments with different data or slightly different models. For machine learning research, reproducibility is complicated highly from the presence of a very large number of stochastic elements such as weight initialisation, which can be ameliorated by \textit{training} fewer weights. This fundamentally encourages foundation models not because they are smaller (often, it is the opposite case) but because for any individual researcher, the number of additional weights needed can be made smaller through using simpler adaptation techniques. It is clear that model reuse is an instrumental value for this type of reproducibility as this facilitated reproduction relies on the capability to directly reuse a foundation model without retraining or otherwise modifying its interior weights. Given its prevalence in the philosophy of science for several decades, it has permeated enough into the general consciousness and, for many, is a fundamental value in itself.

\subsection*{Research speed and facility}
An often overlooked aspect of medical imaging research is that it is difficult. It often requires collecting data which requires strict ethical and security requirements while also developing highly performing solutions to problems in two- or three-dimensional computer vision. Foundation models address these two issues directly through improved data efficiency (thus limiting the cost and effort for data collection) and model reuse (limiting the technological effort needed to reach higher performance) and thus the use foundation models can be motivated simply by improved \textit{research speed and facility}. 

\subsection*{Clinical trust}
\textit{Clinical trust} often relies on more than just the accuracy of the model but also an understanding of how it works especially in concert with other potentially non-independent methods, often relying on the notion of explainability \cite{lin2024machine}. Notions of explainations are highly important in fields such as medicine where the fundamental level of uncertainty and the importance of the decisions being made mean that making the right decision is a matter of having the right justification and following the right guidelines more than the unattainable goal of always being correct \cite{masic2022medical}. Because of this reliance on explainability and justification, clinical trust is therefore an easier goal to achieve with the more transparent methods that can be individually developed rather than from the more direct use of foundation models.

\subsection*{Knowledge extraction}
Also immediately related to explainability and of extreme interest in the use of medical imaging as a scientific research tool rather than a clinical one is the capability to \textit{extract scientific knowledge} from a trained model. That is, if a model gives consistent explanations for the same classes of patients for the same disease, it is more likely that said explanations are an inherent reflection of the disease's etiology. This is especially clear if these explanations are consistent across multiple independent models \cite{estudillo2022voxel} or correlate with other, known comorbidities or demographic variables \cite{estudillo2024non}. Because of the reliance of knowledge extraction on the capability to readily understand the functioning of one or more models, it is inherently more motivating for the creation of one's own models over the use of foundation models. One may argue that this is a question of the scope of knowledge, that foundation models would be able to extract scientific knowledge at a more general level that cuts across domains, although the research that has been performed regarding this for large language models in Newtonian physics currently suggests otherwise \cite{vafa2025has} and this is for domains that are far less complex than human pathologies viewed indirectly through the physical processes underlying medical imaging.

\subsection*{Model recognition}
Of the potential benefits of model reuse, especially a model that is well known in the community, is \textit{modal recognition}, that the name of the model could act as a keyword (even if informally) and that the technological details can be explained with a citation. Again, this value (at least for the people using rather than creating the foundation model) is clearly instrumental. Clearly, the dependence of model recognition on model reuse means that it more clearly encourages the more direct use of foundation models. As with many values, model recognition is again purely instrumental and it is difficult to imagine anyone ending their line of values-based reasoning here without justifying why model recognition is a good thing for other, more fundamental reasons.

\subsection*{Research sovereignty}
One concern in terms of fairness in research more generally is \textit{research sovereignty} or the diversification of researchers themselves with autonomy in terms of defining their own research problems and methodologies. From a geographical lens, it is especially important to increase the capacity for meaningful research in low- and middle-income countries. In these contexts, research facility is of particular importance because if research is not ``easy'' to perform (from the perspective of higher-income countries) it might not be performable at all. Making foundation models more readily available and open-source allows these researchers to engage in their own independent, successful research in a way that cannot be easily offset by concerns of making models lightweight in that the elements actually trained by the researchers themselves can be made more lightweight. Determining exactly how this global research is structured is nuanced and doing so unreflectively can lead to issues such as data colonialism or a lack of respect for local autonomy in terms of problem and solution definition \cite{stanley2024assessing}. There is also a big question for how these models should be housed and distributed as to not encourage dependence on a specific higher-income country which hosts said models (a specific case of ``infrastructure lock-in'' at a global level) \cite{patel2026data,klotz2026buy}. Nevertheless, foundation models at the moment appear to do more to level the playing field in research than the opposite, although this may not necessarily be the case and illustrates an inherent political dimension to how foundation models themselves should be designed and made available.

\subsection*{Environmentalism}
These final two values are where the crisp delineation between values supporting foundation models or not begins to fail. \textit{Environmentalism} in particular is torn between the choice of privileging concerns regarding the \textit{use} of artificial intelligence and the \textit{creation} of said artificial intelligences. Both have significant costs in terms of hardware \cite{curcio2025evaluating}, energy \cite{pimenow2024challenges}, and water \cite{george2023environmental} consumption. In the case of training time, foundation models have a distinct advantage due to limiting the amount of the model that needs to be changed (i.e. \textit{model reuse}) and lower data requirements (i.e. \textit{data efficiency}). However, due to the size of these models and their relative inefficiency for any particular use, they have significant impact in their use which is only exacerbated by their generality. The potential for more computationally frugal artificial intelligence relies on the smaller, simpler models that are more feasible if trained in a more domain- and task-specific manner. 

\subsection*{Publishability and publication impact}
The last value explored in this article is that of \textit{publishability}, often seen as one of indices of academic success if not its sole goal in the case of the ``publish-or-perish'' perspective. This can have facets related to individual papers (and thus have values such as model recognition, novelty, and knowledge extraction as instrumental values) but also facets related to an individual researcher, notably the frequency of their publications, which clearly has research speed and facility as an instrumental value. Thus, publishability alone should not directly motivate a specific technical decision in a consistent way, but depends highly on exactly how it is translated to specific instrumental values.

\section{Fairness: an ambiguous value}
At this stage, we had exhausted all the values that could be cleanly associated with one side of the spectrum or the other with only fundamental values such as environmentalism or publishability affecting both, but there was one important value that could not be so easily classified: \textit{fairness}.

The fairness of the model predictions could be argued to motivate either side of the spectrum depending on future implementation of foundation models. Given that these models \textit{should be} pretrained using very large datasets (which are more likely to include more diverse representation), one could argue that they would lower the bar for researchers using them to ensure a level of fairness. However, this is not actually the case and most foundation models (in our experience) are pretrained by single centres on single datasets. Even if this hurdle is overcome, there are issues that foundation models may perpetuate biases from their pretraining data, similar to what we have seen in commercial generative models for image generation \cite{saravanan2023exploring}. Ensuring a higher level of fairness in foundation models is still possible, but likely requires additional technical research to combine elements of fairness research, such as adversarial debiasing \cite{zhang2018mitigating} in pretraining-time. Currently, fairness concerns are left to those making use of the foundation models, who generally show that there are consistent subgroup biases in the representations learned by foundation models \cite{mohan2026fairness,zheng2025towards}. At the current time, fairness enhancement and bias mitigation techniques tend to be more effective when more of the model can be influenced by them, but this might not apply for all possible techniques in the future. The ambiguity of fairness in foundation models is therefore something that should be debated and better understood. The fact that it does not cleanly support one side of the spectrum or the other illustrates the potential for aspects such as foundation model regulation to step in and provide minimal fairness requirements and auditing for foundation models across a broad spectrum of applications as well as the individual end-use-case models they support; something that has only recently started to be debated in healthcare regulation forums \cite{wojcik2022foundation}.

\section{Discussion and conclusion}
This article presents an initial exploration into the research values that underlie the choice of whether or not to use a foundation model for a new particular medical image processing problem. As with any exploration, the results may be limited by the scope of experience for the particular authors of this article as well as those of the entire community. Medical imaging foundation models represent a particularly new element in a rapidly evolving field, and many of the values mentioned above depend on current state-of-the-art architectures and infrastructures. Developments in adjacent areas such as federated learning \cite{zhuang2023foundation} could allow for foundation model use and training to occur in lock-step, which could introduce a new list of potential benefits (e.g. reduced data requirements at scale, reduced computational requirements, improved fairness and performance) as well as costs (e.g. security against adversarial attacks or training-time data extraction and increased concerns over data colonialism and geographic fairness). Regardless of exactly what the future holds, it is clear that specific research values do have a clear way of motivating the use of foundation models or abstaining from their use and thus give the scientific community a better and more self-reflective understanding of its own workings and the philosophy of science community a deeper appreciation of how specific technological decisions (beyond those of simply what problems are to be addressed and with what data) have philosophical import even in a field as pragmatic and applied as medical image computing.

\section*{Disclosures of interests}
The authors have no conflicts of interest to disclose.

%
% ---- Bibliography ----
%
% BibTeX users should specify bibliography style 'splncs04'.
% References will then be sorted and formatted in the correct style.
%
% \bibliographystyle{splncs04}
% \bibliography{mybibliography}
%
\bibliographystyle{splncs04}
\bibliography{Paper-0021.bib}

@article{baxter2025exploring,
  title={Exploring the values underlying machine learning research in medical image analysis},
  author={Baxter, John SH and Eagleson, Roy},
  journal={Medical Image Analysis},
  volume={102},
  pages={103494},
  year={2025},
  publisher={Elsevier}
}

@article{duffy2009values,
  title={Values in qualitative and quantitative research},
  author={Duffy, Maureen and Chenail, Ronald J},
  journal={Counseling and values},
  volume={53},
  number={1},
  pages={22--38},
  year={2009},
  publisher={Wiley Online Library}
}

@article{masic2014ethics,
  title={Ethics in medical research and publication},
  author={Masic, Izet and Hodzic, Ajla and Mulic, Smaila},
  journal={International journal of preventive medicine},
  volume={5},
  number={9},
  pages={1073},
  year={2014}
}

@article{xiang2025research,
  title={Research topic choice: Motivations, strategies, and consequences},
  author={Xiang, Sidney and Romero, Daniel and Teplitskiy, Misha},
  journal={Quantitative Science Studies},
  volume={6},
  pages={623--651},
  year={2025},
  publisher={MIT Press 255 Main Street, 9th Floor, Cambridge, Massachusetts 02142, USA~…}
}

@article{khan2025comprehensive,
  title={A comprehensive survey of foundation models in medicine},
  author={Khan, Wasif and Leem, Seowung and See, Kyle B and Wong, Joshua K and Zhang, Shaoting and Fang, Ruogu},
  journal={IEEE Reviews in Biomedical Engineering},
  year={2025},
  publisher={IEEE}
}

@article{sun2025foundation,
  title={A foundation model for enhancing magnetic resonance images and downstream segmentation, registration and diagnostic tasks},
  author={Sun, Yue and Wang, Limei and Li, Gang and Lin, Weili and Wang, Li},
  journal={Nature Biomedical Engineering},
  volume={9},
  number={4},
  pages={521--538},
  year={2025},
  publisher={Nature Publishing Group UK London}
}

@article{blankemeier2024merlin,
  title={Merlin: A vision language foundation model for 3d computed tomography},
  author={Blankemeier, Louis and Cohen, Joseph Paul and Kumar, Ashwin and Van Veen, Dave and Gardezi, Syed Jamal Safdar and Paschali, Magdalini and Chen, Zhihong and Delbrouck, Jean-Benoit and Reis, Eduardo and Truyts, Cesar and others},
  journal={Research Square},
  pages={rs--3},
  year={2024}
}

@article{zhu2024medical,
  title={Medical sam 2: Segment medical images as video via segment anything model 2},
  author={Zhu, Jiayuan and Hamdi, Abdullah and Qi, Yunli and Jin, Yueming and Wu, Junde},
  journal={arXiv preprint arXiv:2408.00874},
  year={2024}
}

@inproceedings{tian2024unigradicon,
  title={unigradicon: A foundation model for medical image registration},
  author={Tian, Lin and Greer, Hastings and Kwitt, Roland and Vialard, Fran{\c{c}}ois-Xavier and San Jos{\'e} Est{\'e}par, Ra{\'u}l and Bouix, Sylvain and Rushmore, Richard and Niethammer, Marc},
  booktitle={International Conference on Medical Image Computing and Computer-Assisted Intervention},
  pages={749--760},
  year={2024},
  organization={Springer}
}

@incollection{rooney2017borderlands,
  title={The borderlands between epistemic and non-epistemic values},
  author={Rooney, Phyllis},
  booktitle={Current controversies in values and science},
  pages={31--45},
  year={2017},
  publisher={Routledge}
}

@article{rezaei2020zero,
  title={Zero-shot learning and its applications from autonomous vehicles to COVID-19 diagnosis: A review},
  author={Rezaei, Mahdi and Shahidi, Mahsa},
  journal={Intelligence-based medicine},
  volume={3},
  pages={100005},
  year={2020},
  publisher={Elsevier}
}

@article{wang2023real,
  title={A real-world dataset and benchmark for foundation model adaptation in medical image classification},
  author={Wang, Dequan and Wang, Xiaosong and Wang, Lilong and Li, Mengzhang and Da, Qian and Liu, Xiaoqiang and Gao, Xiangyu and Shen, Jun and He, Junjun and Shen, Tian and others},
  journal={Scientific Data},
  volume={10},
  number={1},
  pages={574},
  year={2023},
  publisher={Nature Publishing Group UK London}
}

@article{jiang2022delve,
  title={Delve into neural activations: Toward understanding dying neurons},
  author={Jiang, Ziping and Wang, Yunpeng and Li, Chang-Tsun and Angelov, Plamen and Jiang, Richard},
  journal={IEEE Transactions on Artificial Intelligence},
  volume={4},
  number={4},
  pages={959--971},
  year={2022},
  publisher={IEEE}
}

@article{liu2024wisdom,
  title={Wisdom of committee: Distilling from foundation model to specialized application model},
  author={Liu, Zichang and Liu, Qingyun and Li, Yuening and Liu, Liang and Shrivastava, Anshumali and Bi, Shuchao and Hong, Lichan and Chi, Ed H and Zhao, Zhe},
  journal={arXiv preprint arXiv:2402.14035},
  year={2024}
}

@article{selvaraju2020grad,
  title={Grad-CAM: visual explanations from deep networks via gradient-based localization},
  author={Selvaraju, Ramprasaath R and Cogswell, Michael and Das, Abhishek and Vedantam, Ramakrishna and Parikh, Devi and Batra, Dhruv},
  journal={International journal of computer vision},
  volume={128},
  number={2},
  pages={336--359},
  year={2020},
  publisher={Springer}
}

@article{ignatiev2019relating,
  title={On relating explanations and adversarial examples},
  author={Ignatiev, Alexey and Narodytska, Nina and Marques-Silva, Joao},
  journal={Advances in neural information processing systems},
  volume={32},
  year={2019}
}

@article{arbel2026primer,
  title={A primer on Bayesian neural networks: review and debates},
  author={Arbel, Julyan and Pitas, Konstantinos and Vladimirova, Mariia and Fortuin, Vincent},
  journal={Statistical Science},
  volume={41},
  number={2},
  pages={316--353},
  year={2026},
  publisher={Institute of Mathematical Statistics}
}

@article{bai2025evlf,
  title={EVLF-FM: Explainable Vision Language Foundation Model for Medicine},
  author={Bai, Yang and Cheng, Haoran and Zhou, Yang and Zhou, Jun and Thirunavukarasu, Arun and Ke, Yuhe and Yao, Jie and Fukutsu, Kanae and Quek, Chrystie Wan Ning and Hong, Ashley and others},
  journal={arXiv preprint arXiv:2509.24231},
  year={2025}
}

@article{nie2025explainable,
  title={An explainable biomedical foundation model via large-scale concept-enhanced vision-language pre-training},
  author={Nie, Yuxiang and He, Sunan and Bie, Yequan and Wang, Yihui and Chen, Zhixuan and Yang, Shu and Cai, Zhiyuan and Wang, Hongmei and Wang, Xi and Luo, Luyang and others},
  journal={arXiv preprint arXiv:2501.15579},
  year={2025}
}

@article{turpin2023language,
  title={Language models don't always say what they think: Unfaithful explanations in chain-of-thought prompting},
  author={Turpin, Miles and Michael, Julian and Perez, Ethan and Bowman, Samuel},
  journal={Advances in Neural Information Processing Systems},
  volume={36},
  pages={74952--74965},
  year={2023}
}

@article{estudillo2022voxel,
  title={Voxel-based diktiometry: Combining convolutional neural networks with voxel-based analysis and its application in diffusion tensor imaging for Parkinson's disease},
  author={Estudillo-Romero, Alfonso and Haegelen, Claire and Jannin, Pierre and Baxter, John SH},
  journal={Human Brain Mapping},
  volume={43},
  number={16},
  pages={4835--4851},
  year={2022},
  publisher={Wiley Online Library}
}

@article{estudillo2024non,
  title={Non-local diffusion-based biomarkers in patients with cocaine use disorder},
  author={Estudillo-Romero, Alfonso and Migliaccio, Raffaella and Batrancourt, B{\'e}n{\'e}dicte and Jannin, Pierre and Baxter, John SH},
  journal={Neuroimage: Reports},
  volume={4},
  number={2},
  pages={100202},
  year={2024},
  publisher={Elsevier}
}

@article{pimenow2024challenges,
  title={Challenges of artificial intelligence development in the context of energy consumption and impact on climate change},
  author={Pimenow, Sergiusz and Pimenowa, Olena and Prus, Piotr},
  journal={Energies},
  volume={17},
  number={23},
  pages={5965},
  year={2024},
  publisher={MDPI}
}

@article{george2023environmental,
  title={The environmental impact of AI: a case study of water consumption by chat GPT},
  author={George, A Shaji and George, AS Hovan and Martin, AS Gabrio},
  journal={Partners Universal International Innovation Journal},
  volume={1},
  number={2},
  pages={97--104},
  year={2023}
}

@article{curcio2025evaluating,
  title={Evaluating the lifecycle economics of AI: The levelized cost of artificial intelligence (LCOAI)},
  author={Curcio, Eliseo},
  journal={Information Systems},
  pages={102634},
  year={2025},
  publisher={Elsevier}
}

@article{lin2024machine,
  title={Machine learning and human-machine trust in healthcare: A systematic survey},
  author={Lin, Han and Han, Jiatong and Wu, Pingping and Wang, Jiangyan and Tu, Juan and Tang, Hao and Zhu, Liuning},
  journal={CAAI Transactions on Intelligence Technology},
  volume={9},
  number={2},
  pages={286--302},
  year={2024},
  publisher={Wiley Online Library}
}

@article{masic2022medical,
  title={Medical decision making-an overview},
  author={Masic, Izet},
  journal={Acta Informatica Medica},
  volume={30},
  number={3},
  pages={230},
  year={2022}
}

@inproceedings{stanley2024assessing,
  title={Assessing the impact of sociotechnical harms in ai-based medical image analysis},
  author={Stanley, Emma AM and Souza, Raissa and Winder, Anthony J and Wilms, Matthias and Pike, G Bruce and Dagasso, Gabrielle and Nielsen, Christopher and MacEachern, Sarah J and Forkert, Nils D},
  booktitle={MICCAI Workshop on Fairness of AI in Medical Imaging},
  pages={163--175},
  year={2024},
  organization={Springer}
}

@article{vafa2025has,
  title={What has a foundation model found? using inductive bias to probe for world models},
  author={Vafa, Keyon and Chang, Peter G and Rambachan, Ashesh and Mullainathan, Sendhil},
  journal={arXiv preprint arXiv:2507.06952},
  year={2025}
}

@article{zhuang2023foundation,
  title={When foundation models meet federated learning: Motivations, challenges, and future directions},
  author={Zhuang, Weiming and Chen, Chen and Li, Jingtao and Chen, Chaochao and Jin, Yaochu and Lyu, Lingjuan},
  journal={arXiv preprint arXiv:2306.15546},
  year={2023}
}

@article{saravanan2023exploring,
  title={Exploring social bias in downstream applications of text-to-image foundation models},
  author={Saravanan, Adhithya Prakash and Kocielnik, Rafal and Jiang, Roy and Han, Pengrui and Anandkumar, Anima},
  journal={arXiv preprint arXiv:2312.10065},
  year={2023}
}

@inproceedings{zhang2018mitigating,
  title={Mitigating unwanted biases with adversarial learning},
  author={Zhang, Brian Hu and Lemoine, Blake and Mitchell, Margaret},
  booktitle={Proceedings of the 2018 AAAI/ACM Conference on AI, Ethics, and Society},
  pages={335--340},
  year={2018}
}

@inproceedings{mohan2026fairness,
  title={A Fairness Audit of Medical Imaging Foundation Models on a Multimodal Structured Clinical Benchmark},
  author={Mohan, Milan and Lingisetty, Saketh and Ahmad, Umar and Kumar, Avi and Harikrishnan, Shankar and Chamarty, Sanjana and Zhu, Kevin},
  booktitle={ICML 2026 Workshop on Structured Data for Health}
}

@article{zheng2025towards,
  title={Towards fair medical ai: Adversarial debiasing of 3d ct foundation embeddings},
  author={Zheng, Guangyao and Jacobs, Michael A and Braverman, Vladimir and Parekh, Vishwa S},
  journal={arXiv preprint arXiv:2502.04386},
  year={2025}
}

@article{patel2026data,
  title={Data Sovereignty and Algorithmic Dependency in the Global South: A Systematic Review of Foundation Model Governance, Digital Inequity, and Decolonial Artificial Intelligence Frameworks},
  author={Patel, Dharma},
  year={2026}
}

@article{klotz2026buy,
  title={The Buy-or-Build Decision, Revisited: How Agentic AI Changes the Economics of Enterprise Software},
  author={Klotz, David},
  journal={arXiv preprint arXiv:2604.26482},
  year={2026}
}

@article{zhang2026data,
  title={Data-centric foundation models in computational healthcare: A survey},
  author={Zhang, Yunkun and Gao, Jin and Tan, Zheling and Zhou, Lingfeng and Ding, Kexin and Zhou, Mu and Zhang, Shaoting and Wang, Dequan},
  journal={ACM Computing Surveys},
  volume={58},
  number={11},
  pages={1--35},
  year={2026},
  publisher={ACM New York, NY}
}

@article{villalobos2022machine,
  title={Machine learning model sizes and the parameter gap},
  author={Villalobos, Pablo and Sevilla, Jaime and Besiroglu, Tamay and Heim, Lennart and Ho, Anson and Hobbhahn, Marius},
  journal={arXiv preprint arXiv:2207.02852},
  year={2022}
}

@article{girdhar2025comprehensive,
  title={A comprehensive review of frugal artificial intelligence: challenges, applications, and the road to sustainable AI},
  author={Girdhar, Nancy and Raj, Aditya and Sharma, Deepak and Singh, Vinay and Doucet, Antoine and Renz, Matthias},
  journal={Soft Computing},
  volume={29},
  number={13},
  pages={4823--4856},
  year={2025},
  publisher={Springer}
}

@inproceedings{wojcik2022foundation,
  title={Foundation models in healthcare: opportunities, biases and regulatory prospects in Europe},
  author={W{\'o}jcik, Malwina Anna},
  booktitle={International Conference on Electronic Government and the Information Systems Perspective},
  pages={32--46},
  year={2022},
  organization={Springer}
}
\end{document}